\documentclass[twocolumn]{aastex63}
\usepackage[T1]{fontenc}
\usepackage{ae,aecompl}
\usepackage{graphicx}
\usepackage{amsmath}
\usepackage{amssymb}
\usepackage{supertabular}
\usepackage{color}
\usepackage{longtable}
\usepackage{booktabs}
\usepackage{textcomp}

\shorttitle{BH growths in GRBs for the low mass gap}
\shortauthors{Qu \& Liu}

\begin{document}

\title{Revisiting black hole hyperaccretion in the center of gamma-ray bursts for the lower mass gap}

\author{Hui-Min Qu}
\affiliation{Department of Astronomy, Xiamen University, Xiamen, Fujian 361005, China}

\author[0000-0001-8678-6291]{Tong Liu}
\affiliation{Department of Astronomy, Xiamen University, Xiamen, Fujian 361005, China}

\correspondingauthor{Tong Liu}
\email{tongliu@xmu.edu.cn}

\begin{abstract}
The ultrarelativistic jets triggered by neutrino annihilation processes or Blandford-Znajek (BZ) mechanisms in stellar-mass black hole (BH) hyperaccretion systems are generally considered to power gamma-ray bursts (GRBs). Due to the high accretion rate, the central BHs might grow rapidly on a short timescale, providing a new way to understand ``the lower mass gap'' problem. In this paper, we use the BH hyperaccretion model to investigate BH mass growth based on observational GRB data. The results show that (i) if the initial BH mass is set as $3~M_\odot$, the neutrino annihilation processes are capable of fueling the BHs to escape the lower mass gap for more than half of long-duration GRBs (LGRBs), while the BZ mechanism is inefficient on triggering BH growths for LGRBs; (ii) the mean BH mass growths in the case of LGRBs without observable supernova (SN) association are much larger than these in the case of LGRBs associated with SNe for both mechanisms, which imply that more massive progenitors or lower SN explosion energies prevail throughout the former cases; (iii) for the short-duration GRBs, the mean BH mass growths are satisfied with the mass supply limitation in the scenario of compact object mergers, but the hyperaccretion processes are unable to rescue BHs from the gap in binary neutron star (NS) mergers or the initial BH mass being $3~M_\odot$ after NS-BH mergers.
\end{abstract}


\section{Introduction} \label{sec:intro}

Gamma-ray bursts (GRBs) are the most luminous explosions in the universe. According to their durations, GRBs can be classified into two categories: short-duration GRBs (SGRBs; $T_{90}<2 \mathrm{~s}$) and long-duration GRBs \citep[LGRBs; $T_{90}>2 \mathrm{~s}$, see][]{1993ApJ...413L.101K}. SGRBs are generally believed to be produced by merger events of two compact objects, i.e., two neutron stars (NSs) or an NS and a black hole \citep[BH, e.g.,][]{1989Natur.340..126E,1992ApJ...395L..83N,2007PhR...442..166N}, and LGRBs are widely considered to originate from the collapse of massive stars \citep[e.g.,][]{1993ApJ...405..273W,2006ARA&A..44..507W,2012ARNPS..62..407J}. Moreover, some LGRBs are associated with Type Ib/c supernovae \citep[SNe, see e.g.,][]{2003Natur.423..847H,2004ApJ...609L...5M,2011ApJ...743..204B,2012grb..book..169H,2015Natur.523..189G}, which sheds light on the progenitors and central engines of LGRBs.

Two popular models have been proposed for the central engines of GRBs, involving a rotating stellar-mass BH surrounded by a hyperaccretion disk \citep[e.g.,][]{1991AcA....41..257P,1992ApJ...395L..83N,1999ApJ...524..262M,2017NewAR..79....1L} and a millisecond magnetar \citep[e.g.,][]{1992ApJ...392L...9D,1992Natur.357..472U,1998A&A...333L..87D,1998PhRvL..81.4301D,1998ApJ...505L.113K,2001ApJ...552L..35Z,2011MNRAS.413.2031M}. The GRB jets could be powered either by the rotational energy of the magnetars or by the gravitational or rotational energy of the accreting BHs. In the BH hyperaccretion scenario, neutrinos radiated from the heated disk matter can liberate the gravitational energy and then annihilate outside of the disk to produce GRB jets \citep[e.g.,][]{1997A&A...319..122R,2003MNRAS.345.1077R,2011MNRAS.410.2302Z}. This hyperaccretion mode is called neutrino-dominated accretion flows \citep[NDAF, e.g.,][]{1999ApJ...518..356P,2001ApJ...557..949N,2004MNRAS.355..950J,2005ApJ...629..341K,2005ApJ...632..421L,2006ApJ...643L..87G,2007ApJ...657..383C,2007ApJ...662.1156K,2007ApJ...661.1025L,2009ApJ...700.1970L,2013ApJS..207...23X}. For a recent review see \citet{2017NewAR..79....1L}. Alternatively, the strong magnetic fields threading the BH horizon can also power the Poynting jets to efficiently extract the BH's rotational energy, namely, the Blandford-Znajek (BZ) mechanism \citep{1977MNRAS.179..433B,2000PhR...325...83L,2003MNRAS.345.1077R}. In our work, we applied the neutrino annihilation process and BZ mechanism to investigate BH mass growth, and it is interesting to note that neutrino annihilation, as the initial dominant mechanism, could be replaced by BZ jets when the accretion rate decreases \citep[e.g.,][]{2017ApJ...850...30L,2018ApJ...852...20L}.

In the hyperaccretion system, the BH mass and spin should undergo drastic evolution \citep[e.g.,][]{2015ApJ...806...58L,2015ApJ...815...54S}. According to the GRB progenitor models, the initial BH mass is generally considered to approximately 3 $M_\odot$. Thus, the BH mass growth in the center of GRBs should be related to the lower mass gap (or the first mass gap) in the mass distribution of the compact objects. This gap (very few compact objects exist in the range of $\sim 2-5$ or $2.5-5$ or $3-5~ M_{\odot}$) was discovered in the statistical analyses of the X-ray binary observations \citep{2010ApJ...725.1918O,2011ApJ...741..103F}.

Three $\sim 2~M_\odot$ NSs were measured using the ``Shapiro delay'' effects \citep{2010Natur.467.1081D,2013Sci...340..448A,2020NatAs...4...72C}. Recently, NASA's \emph{NICER} constrained the mass measure of PSR J0740+6620, $2.072^{+0.067}_{-0.066}~M_\odot$ \citep{2021ApJ...918L..28M,2021ApJ...918L..27R}. Furthermore, \citet{2019Sci...366..637T} reported a $\sim 3 ~M_\odot$ BH candidate in a noninteracting low-mass binary system. In the aLIGO/Virgo detections, the compact remnants of GWs 170817 \citep{2017PhRvL.119p1101A} and 190425 \citep{2020ApJ...892L...3A} and one of the objects participating in GW 190814 \citep{2020ApJ...896L..44A} are all in the gap. One can find that a lower mass gap exists, but a small number of compact objects still remain here. \citet{2012ApJ...757...91B} proposed that the rapid explosion mechanism of core-collapse SNe (CCSNe) could absorb the newborn remnants from the gap. \citet{2021ApJ...908..106L} simulated that the gap can be naturally built by the low explosion energy dominated distribution of CCSNe.

In this paper, by using a GRB sample, we revisit the BH hyperaccretion systems with a neutrino annihilation process and BZ mechanism and then analyze the effects of BH mass growth on the lower mass gap. This paper is organized as follows. In Section 2, we present the analytical models for describing the evolution of a Kerr BH and estimating the BH mass growth. The main results are shown in Section 3. Conclusions and discussions are made in Section 4.

\section{Model} \label{sec:model}
\subsection{BH evolution}

As a plausible central engine of GRBs, a rotating stellar BH surrounded by a hyperaccretion disk with a very high accretion rate should trigger violent evolution of BH characteristics. Based on the conservation of energy and angular momentum, the mass and angular momentum of the BH evolve with time as \citep[e.g.,][]{2012ApJ...760...63L}
\begin{equation}
\frac{d M_{\mathrm{BH}}}{d t}=\dot{M} e_{\mathrm{ms}},
\label{eq1}
\end{equation}
\begin{equation}
\frac{d J_{\mathrm{BH}}}{d t}=\dot{M} l_{\mathrm{ms}},
\label{eq2}
\end{equation}
where $M_{\rm BH}$ and $J_{\rm BH}$ are the mass and angular momentum of the BH, $\dot{M}$ is the mass accretion rate, and $e_{\mathrm{ms}}$ and $l_{\mathrm{ms}}$ are the specific energy and angular momentum corresponding to the marginally stable orbit radius of the BH. They are defined as $e_{\mathrm{ms}}=\frac{1}{\sqrt{3 x_{\mathrm{ms}}}}\left(4-\frac{3 a_{*}}{\sqrt{x_{\mathrm{ms}}}}\right)$ and $l_{\mathrm{ms}}=2 \sqrt{3} \frac{G M_{\mathrm{BH}}}{c}\left(1-\frac{2 a_{*}}{3 \sqrt{x_{\mathrm{ms}}}}\right)$, respectively, where $a_{*} \equiv c J_{\mathrm{BH}} / G M_{\mathrm{BH}}^{2}$ ($0\leq a_{*}\leq 1$) is the dimensionless spin parameter of the BH and $x_{\rm{ms}}$ is the dimensionless marginally stable orbit radius of the disk, which is defined as $x_{\mathrm{ms}}=3+Z_{2}-\sqrt{(3-Z_{1})(3+Z_{1}+2 Z_{2})}$ with $Z_{1}=1+(1-a_{*}^{2})^{1/3}[(1+a_{*})^{1/3}+(1-a_{*})^{1/3}]$ and $Z_{2}=\sqrt{3 a_{*}^{2}+Z_{1}^{2}}$ \citep[e.g.,][]{1972ApJ...178..347B,1998GrCo....4S.135N,2008bhad.book.....K}.

By combining Equations (1) and (2), the evolution of the BH spin can be expressed by \citep[e.g.,][]{2014ApJ...781L..19H}
\begin{equation}
\frac{d a_{*}}{d t}=2 \sqrt{3} \frac{\dot{M}}{M_{\mathrm{BH}}}\left(1-\frac{a_{*}}{\sqrt{x_{\mathrm{ms}}}}\right)^{2}.
\label{eq3}
\end{equation}

For the BZ mechanism, a part of the BH rotational energy will be extracted by the Poynting jet, which would affect the evolution of the BH mass and angular momentum as \citep[e.g.,][]{2000PhR...325...83L,2000JKPS...36..188L}
\begin{equation}
	\frac{d M_{\mathrm{BH}}}{d t}=\dot{M} e_{\mathrm{ms}}-\frac{L_{\mathrm{BZ}}}{c^2},
	\label{eq4}
\end{equation}
\begin{equation}
	\frac{d J_{\mathrm{BH}}}{d t}=\dot{M} l_{\mathrm{ms}}-\frac{L_{\mathrm{BZ}}}{c^2 \Omega_{\mathrm{F}}},
	\label{eq5}
\end{equation}
where $L_{\mathrm{BZ}}$ is the BZ jet power and $\Omega_{\mathrm{F}}$ is the magnetic field angular velocity at the marginally stable orbit radius. We adopt the optimal mode $\Omega_{\mathrm{F}}=\Omega_{\mathrm{H}} / 2$ here \citep[e.g.,][]{2000PhR...325...83L,2000JKPS...36..188L}, where $\Omega_{\mathrm{H}} \equiv a_{*} c^3 / [2 (1+\sqrt{1-a_{*}^2}) GM_{\mathrm{BH}}]$ is the angular velocity on the stretched horizon. As the estimations, for the BZ jet power ranging from $10^{49}$ to $10^{50} ~\mathrm{erg} ~\mathrm{s}^{-1}$ and $\Omega_{\mathrm{F}} \sim 10^{4}~ \mathrm{s}^{-1}$, the fraction of the angular momentum extracted, $L_{\mathrm{BZ}}/c^2 \Omega_{\mathrm{F}}$, would be negligible, and the extracted rest-mass energy is relatively small as well. For the BZ luminosity up to $\sim 10^{51} ~\mathrm{erg} ~\mathrm{s}^{-1}$ and lasting time $\sim 50 ~\mathrm{s}$, the fraction of the mass extracted is about 1$\%$. Nevertheless, we take these effects into account in the below calculations for the BZ mechanism.

According to the above equations, we can obtain the time-dependent characteristics of the BH once the initial mass $M_{\rm BH,0}$ and spin $a_{*,0}$ of the BH and the BZ jet power are given.

\subsection{Two mechanisms}

The mean luminosity of the GRB jet can be estimated as \citep[e.g.,][]{2011ApJ...739...47F,2015ApJ...806...58L}
\begin{equation}
L_{\mathrm{j}} \simeq \frac{\left(E_{\gamma, \mathrm{iso}}+E_{\mathrm{k}, \mathrm{iso}}\right)(1+z) \theta_{\mathrm{j}}^{2}}{2 T_{90}},
\label{eq6}
\end{equation}
where $E_{\gamma, \mathrm{iso}}$ is the isotropic radiated energy in the prompt emission phase, $E_{\mathrm{k}, \mathrm{iso}}$ is the isotropic kinetic energy powering long-lasting afterglow, $z$ is the redshift, $\theta_{\mathrm{j}}$ is the half-opening angle of the jet, and $T_{90}$ can be roughly considered as the duration of the violent activity of the central engine. Note that we take the time-independent jet luminosity of GRBs, thus the accretion rate $\dot{M}$ would be time-dependent in the BH evolution.

For a BH hyperaccretion system in the center of GRBs, the energy output given by Equation (\ref{eq6}) is determined by the neutrino annihilation luminosity $L_{\nu \bar{\nu}}$ or the BZ jet power $L_{\mathrm{BZ}}$. The annihilation luminosity can be written as a function of the BH mass accretion rate and the spin parameter $a_{*}$ \citep{2011MNRAS.410.2302Z}, i.e.,
\begin{eqnarray}
L_{\nu\bar{\nu}} && \approx 1.59 \times 10^{54} ~x_{\rm ms}^{-4.8}~m_{_{\rm BH}}^{-3/2} \nonumber \\&&
\times \Bigg \{
\begin{array}{ll}
0 & \hbox{for $\dot{m} < \dot {m}_{\rm ign}$}\\
\dot{m }^{9/4} & \hbox{for $\dot {m}_{\rm ign} < \dot{m} < \dot{m}_{\rm trap}$}\\
\dot{m}_{\rm trap}^{9/4} & \hbox{for $\dot{m} > \dot{m}_{\rm trap}$}\\
\end{array}
\Bigg \} ~\rm erg~s^{-1},
\label{eq7}
\end{eqnarray}
where $m_{\rm BH}=M_{\rm BH}/M_\odot$, $\dot{m}=\dot{M}/(M_\odot ~\rm s^{-1})$, $\dot{m}_{\text {ign}}$ is the dimensionless critical ignition accretion rate, $\sim 0.001~M_\odot~\rm s^{-1}$, and $\dot{m}_{\text {trap}}$ are the dimensionless accretion rates if neutrino trapping appears \citep[e.g.,][]{2007ApJ...657..383C,2013ApJS..207...23X,2015ApJ...815...54S}.

The BZ jet power can be estimated by \citep[e.g.,][]{2018ApJ...852...20L}
\begin{equation}
L_{\mathrm{BZ}}=9.3 \times 10^{53} a_{*}^{2} \dot{m} X\left(a_{*}\right)~\mathrm{erg} ~\mathrm{s}^{-1},
\label{eq8}
\end{equation}
and
\begin{equation}
X\left(a_{*}\right)=F\left(a_{*}\right) /(1+\sqrt{1-a_{*}^{2}})^{2},
\end{equation}
where $F\left(a_{*}\right)=\left[\left(1+q^{2}\right) / q^{2}\right][(q+1 / q) \arctan (q)-1]$ with $q=a_{*} /(1+\sqrt{1-a_{*}^{2}})$.

In our simple calculations, we first obtain the time-independent GRB jet luminosities using observational data based on Equation (\ref{eq6}). By applying two different mechanisms, the mass accretion rate at each time step can be obtained as the function of $L_{\mathrm{j}}$:
\begin{equation}
\dot{m}_{\nu\bar{\nu}} = (\frac{L_{\mathrm{j}}}{1.59 \times 10^{54} x_{\rm ms}^{-4.8} m_{_{\rm BH}}^{-3/2} ~\rm erg~s^{-1}})^{4/9},
\label{eq10}
\end{equation}
or
\begin{equation}
\dot{m}_{\mathrm{BZ}} = \frac{L_{\mathrm{j}}}{9.3 \times 10^{53} a_{*}^{2} X\left(a_{*}\right)~\mathrm{erg} ~\mathrm{s}^{-1}}.
\label{eq11}
\end{equation}

Incorporating the values of $\dot{m}$, $m_{\rm BH}$, and $a_{*}$ at last time step into BH evolution functions (Equations (\ref{eq1}) and (\ref{eq2}) or (\ref{eq4}) and (\ref{eq5})), the mass and spin of the BH at next time step can be solved until the time reaches $T_{90}$, then the final BH masses $M_{\rm BH,f}$ are obtained. The main results are discussed below.

\section{Results} \label{sec:results}

We adopt the data of 14 LGRBs associated with SNe \citep[hereafter LGRBs-SNe,][]{2019ApJ...871..117S}, 40 LGRBs without observable SN association \citep[hereafter LGRBs-noSNe,][]{2017JHEAp..13....1Y}, and 31 SGRBs \citep{2015ApJ...806...58L} to calculate the BH mass growth in the BH hyperaccretion systems with neutrino annihilation processes and the BZ mechanism. The durations, redshifts, half-opening angles, $E_{\gamma, \mathrm{iso}}$, and $E_{\mathrm{k}, \mathrm{iso}}$ are included. In the LGRB-noSN sample, the redshifts are in the range of $\sim 0.542-4.394$. The absence of SNe in the LGRB-noSN cases does not certainly mean the failures of SN explosions, but may be the results of explosions being too weak or too distant.

\subsection{Initial BH mass}

\begin{figure}
\centering
\includegraphics[width=1.1\linewidth]{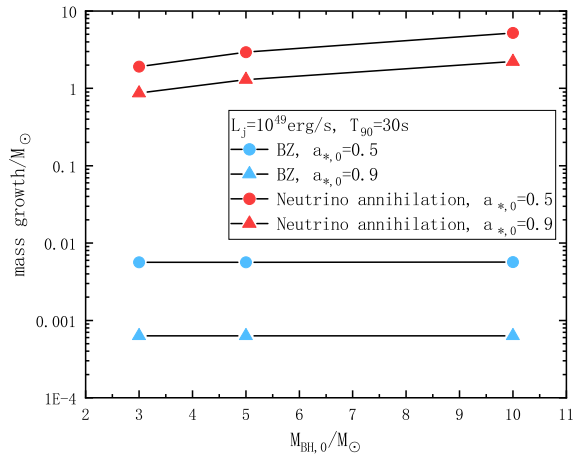}
\caption{Influence of initial BH masses on BH mass growths in a typical GRB case with the luminosity $L_{\mathrm{j}}=10^{49} \mathrm{erg} ~\mathrm{s}^{-1}$ and duration $T_{90}=30s$ for two mechanisms and different initial BH spins. The circles and triangles represent $a_{*,0}$ = 0.5 and 0.9, in which the blue and red colors denote the BZ mechanism and the neutrino annihilation process, respectively.}
\label{fig1}
\end{figure}

For the merger scenario, according to aLIGO/Virgo detections, the remnant mass after merger before accretion in GW 170817 \citep{2017PhRvL.119p1101A} is close to $3~M_{\odot}$. Recently, two sources of the NS-BH coalescence, GWs 200105 and 200115, have the initial BH component masses $5.7_{-2.1}^{+1.8}\ M_{\odot}$ and $8.9_{-1.5}^{+1.2}\ M_{\odot}$ for high spin case before mergers although there in no observably associated electromagnetic counterparts after mergers \citep{2021ApJ...915L...5A}. For the scenario of a CCSN with a progenitor star in the range of $20-40~M_\odot$, an iron core with mass around $2.1-2.4 ~M_{\odot}$ is produced \citep[e.g.,][and references therein]{2008ApJS..174..223B,2021ApJ...908..106L}, then its mass will reach or exceed 3 $M_{\odot}$ due to the hyperaccretion process within several seconds for the initial explosion energy lower than $4 \times 10^{51} ~\rm erg$. Thus, when the jets break out from the envelope and trigger an observable GRB, the accreting BH is likely to have an initial mass around or more than 3 $M_{\odot}$ \citep[e.g.,][]{2021MNRAS.507..431W}.

Before the discussion for all collected GRB cases, we should test the effects of the initial BH mass on the BH mass growth in our model. In Figure 1, we adopt the typical GRB luminosity $L_{\mathrm{j}}=10^{49} ~\mathrm{erg} ~\mathrm{s}^{-1}$ and duration $T_{90}=30~\rm s$ to calculate the BH mass growths for the different initial BH masses, i.e., 3, 5, and 10 $M_\odot$. The initial BH mass almost has no impact on the BH mass growths for $a_{*,0}$ = 0.5 and 0.9 in the scenario of the BZ mechanism, while there would be positive correlation between the initial BH mass and the mass growth if the jets are entirely powered by the neutrino annihilation process. The reason is that the larger initial BH mass leads to the larger inner radius of the disk, then the lower temperature at the inner region and lower neutrino luminosity for a certain accretion rate. Of course, from Equation (\ref{eq10}), one can see that a more massive BH would have a larger accretion rate for a given neutrino annihilation luminosity, thus leading to the more effective mass growth. For the BZ mechanism, the initial BH mass has no affect on the accretion rate, as shown in Equation (\ref{eq11}). Besides, the initial BH spin would also affect mass growths by altering the dimensionless inner stable orbit. By considering the above theoretical and observational results, $M_{\rm BH,0}=3~M_{\odot}$ and $a_{*,0}$ = 0.5 and 0.9 are adopted in the below calculations.

\subsection{LGRB-SN case}

In Figure 2, we demonstrate the final BH mass $M_{\rm BH,f}$ distributions for LGRB-SN, LGRB-noSN, and SGRB cases. The red and blue bars correspond to the neutrino annihilation process and the BZ mechanism, respectively. The dark and light colors denote the initial BH spin $a_{*,0}$ = 0.5 and 0.9, respectively. Since the accretion rate for the BZ mechanism can be significantly lower than that for the neutrino annihilation mechanism for the same output energy, it can be seen that the mass growth under the BZ mechanism is less efficient than the neutrino annihilation process for all GRB cases. In other words, the neutrino annihilation mechanism would be an easier way for a BH to escape the lower mass gap, especially for the long accretion timescale. Moreover, one can expect that a smaller initial BH spin parameter is favored for BH mass growth. Obviously, the mean jet luminosity is weaker for both lower accretion rates and lower BH spin values, as shown in Equations (\ref{eq7}) and (\ref{eq8}).

The physical relation between LGRBs and SNe is firmly established with the accumulated evidence \citep[e.g.,][]{2003Natur.423..847H,2009ApJ...703.1696Z}, and it has been widely accepted that these LGRB-SN events are born out of the deaths of massive stars ($>8~ M_{\odot}$). In the collapse phase of $\sim 20-40~M_\odot$ progenitors, the core inevitably collapses to form a proto-NS and then continues to collapse into a BH, creating a large amount of neutrinos. Neutrino irradiation revives the stalled shock launched at the core bounce and pushes off the remainder of the star, powering a CCSN \citep[e.g.,][and references therein]{1992A&A...264..105M,1995ApJS..101..181W,2008ApJ...679..639Z,2012ApJ...749...91F,2021ApJ...908..106L}. Then, the fallback hyperaccretion on the central BH powers the ultrarelativistic jets. Once the jets break out from the envelope in the line-of-sight direction, a GRB can be observed. For the more massive progenitor stars ($>40~M_\odot$), the core directly collapses to form BHs and is generally larger than approximately $5~M_\odot$ \citep[e.g.,][]{2002ApJ...567..532H}. Thus, we set $M_{\rm BH,0}=3~M_\odot$ to analyze the final BH mass distribution using the LGRB-SN sample.

\begin{figure*}
	\includegraphics[width=0.5\linewidth]{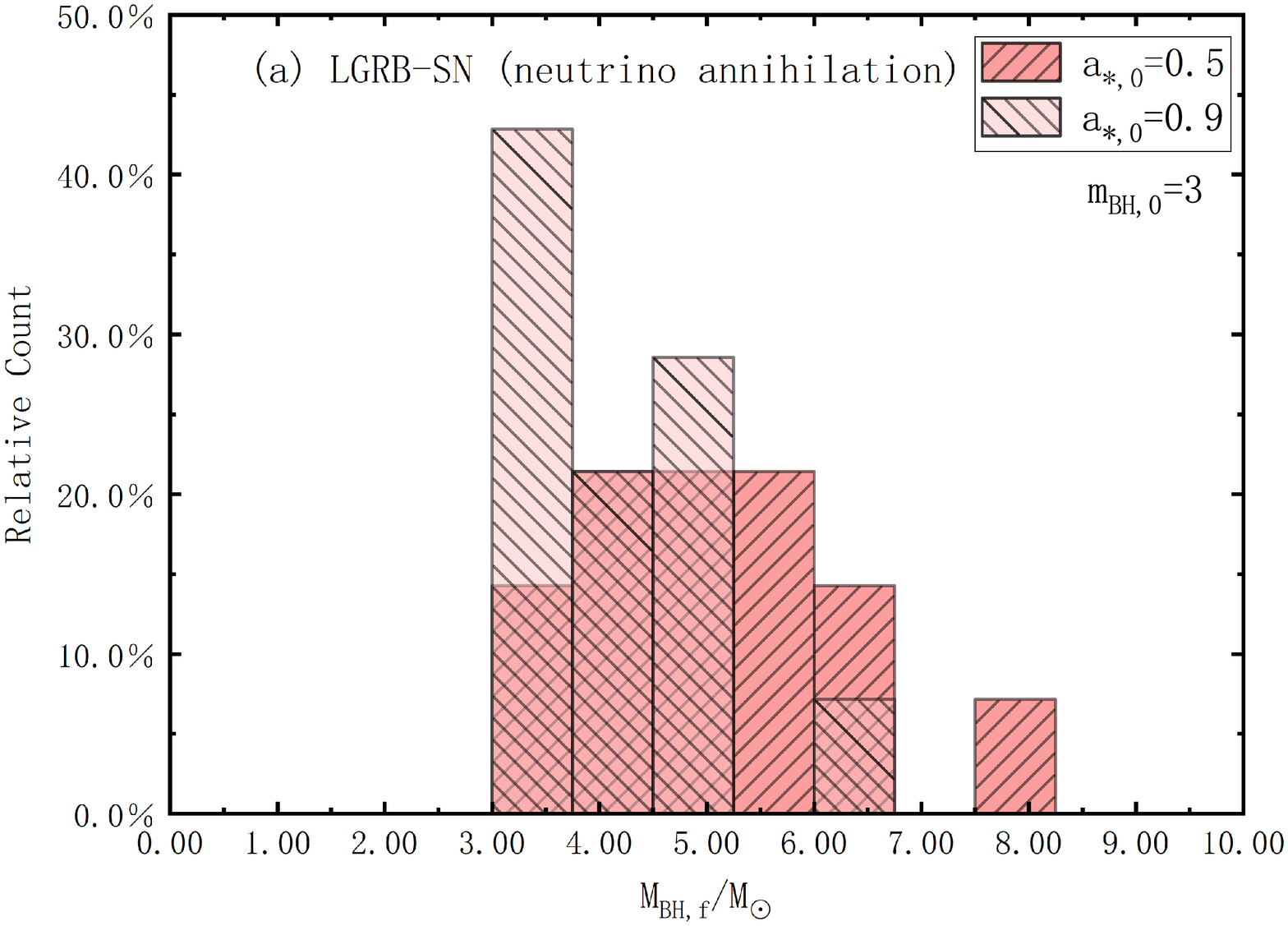}
	\includegraphics[width=0.5\linewidth]{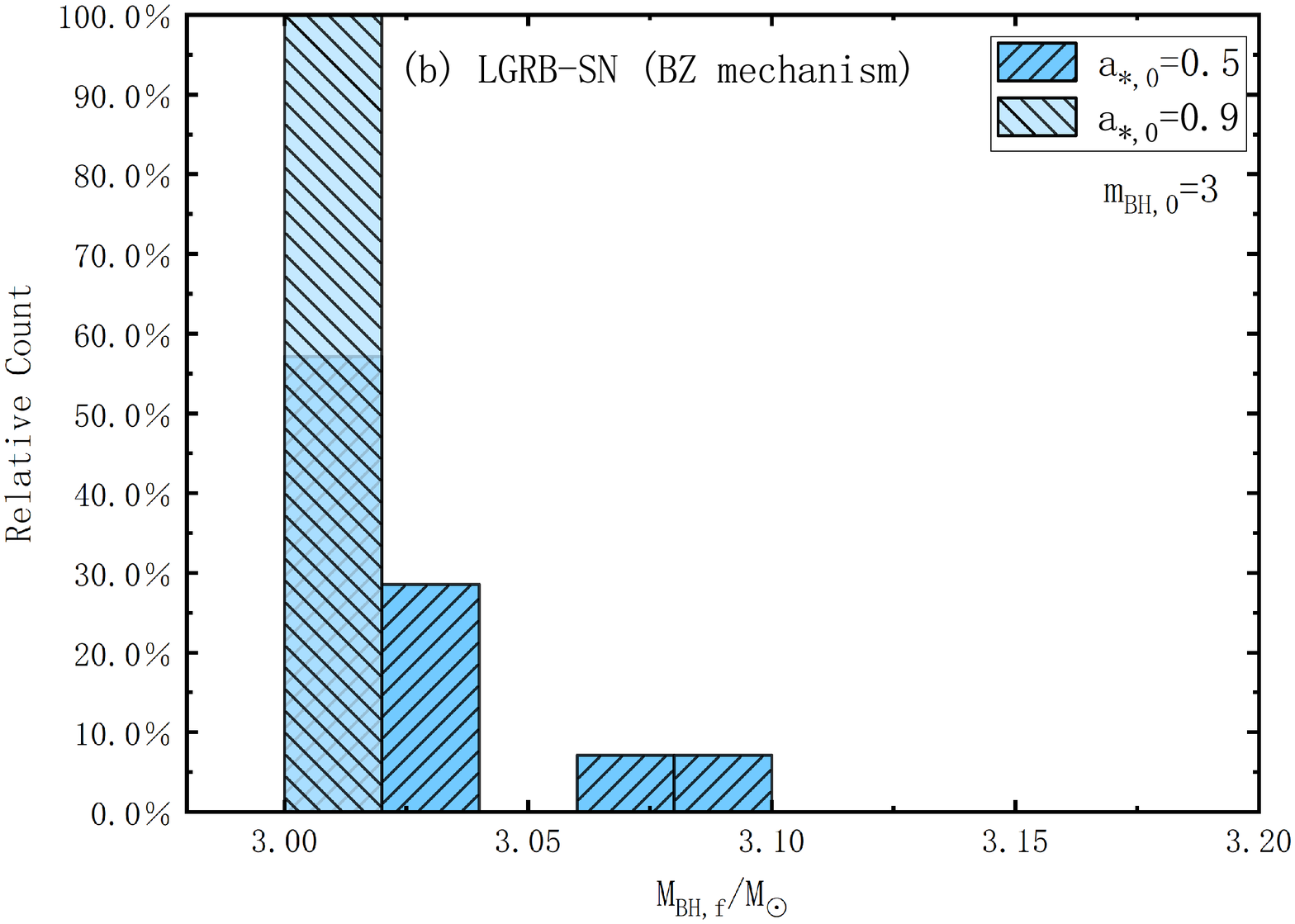}
	\includegraphics[width=0.5\linewidth]{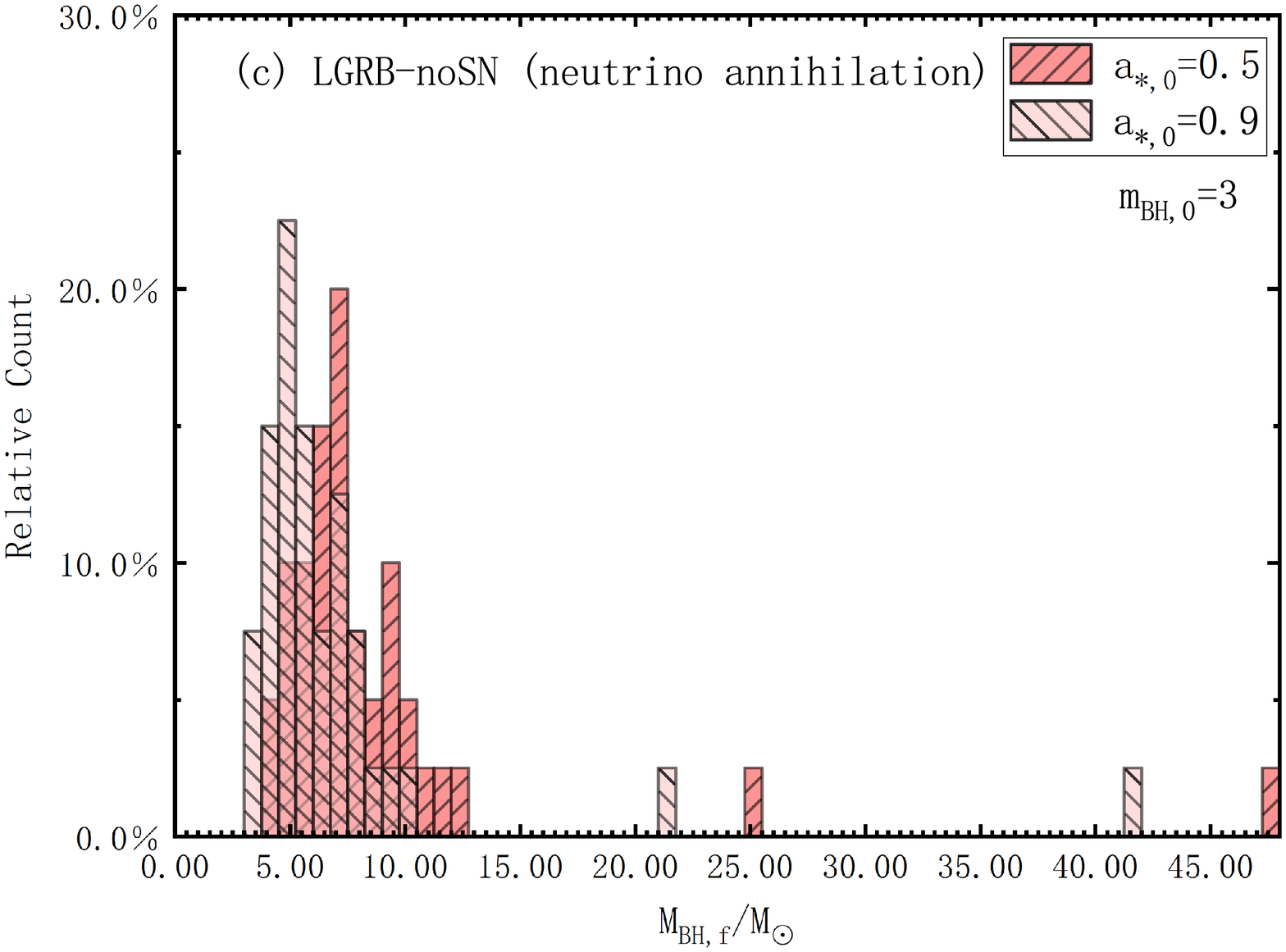}
	\includegraphics[width=0.5\linewidth]{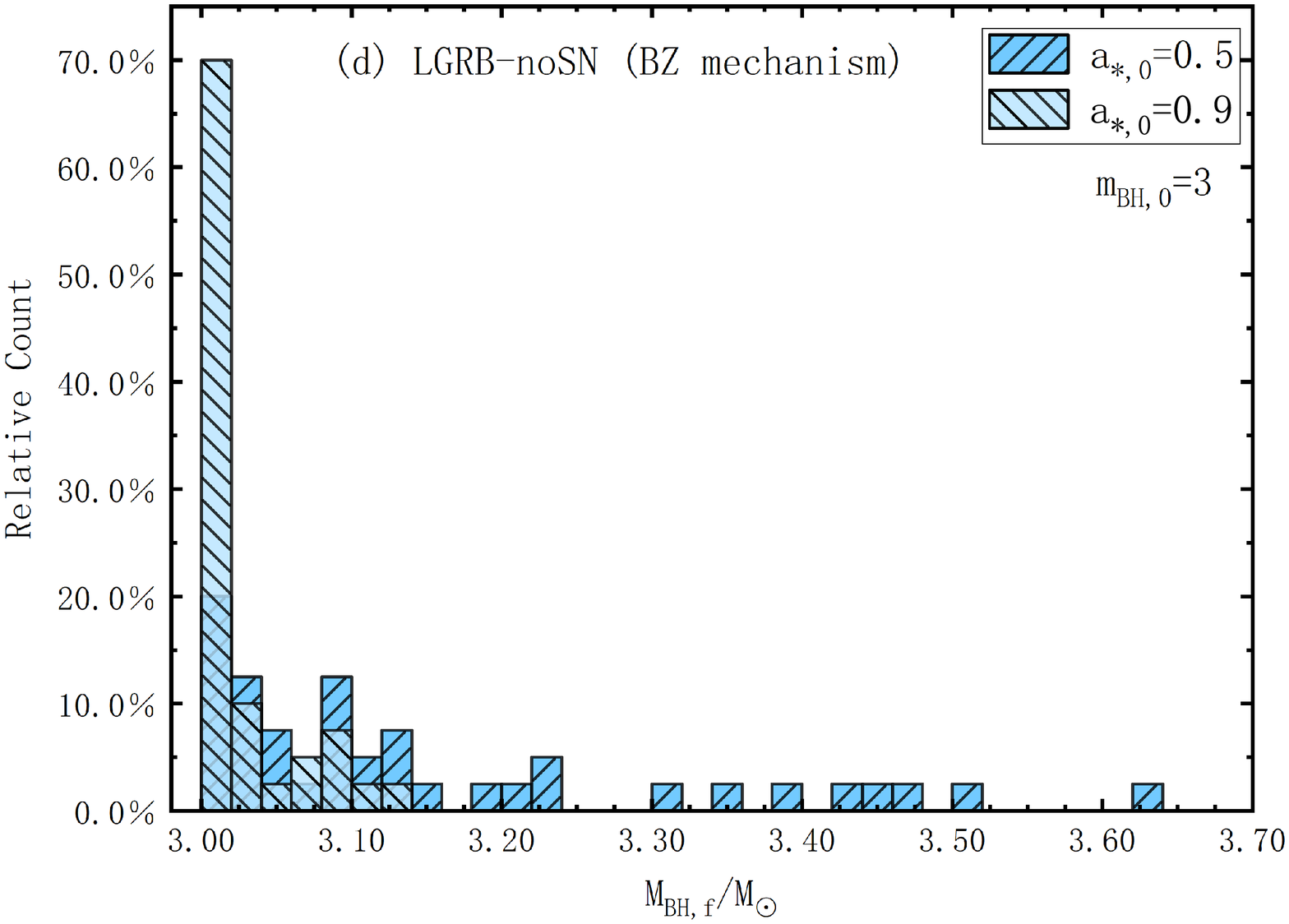}
	\includegraphics[width=0.5\linewidth]{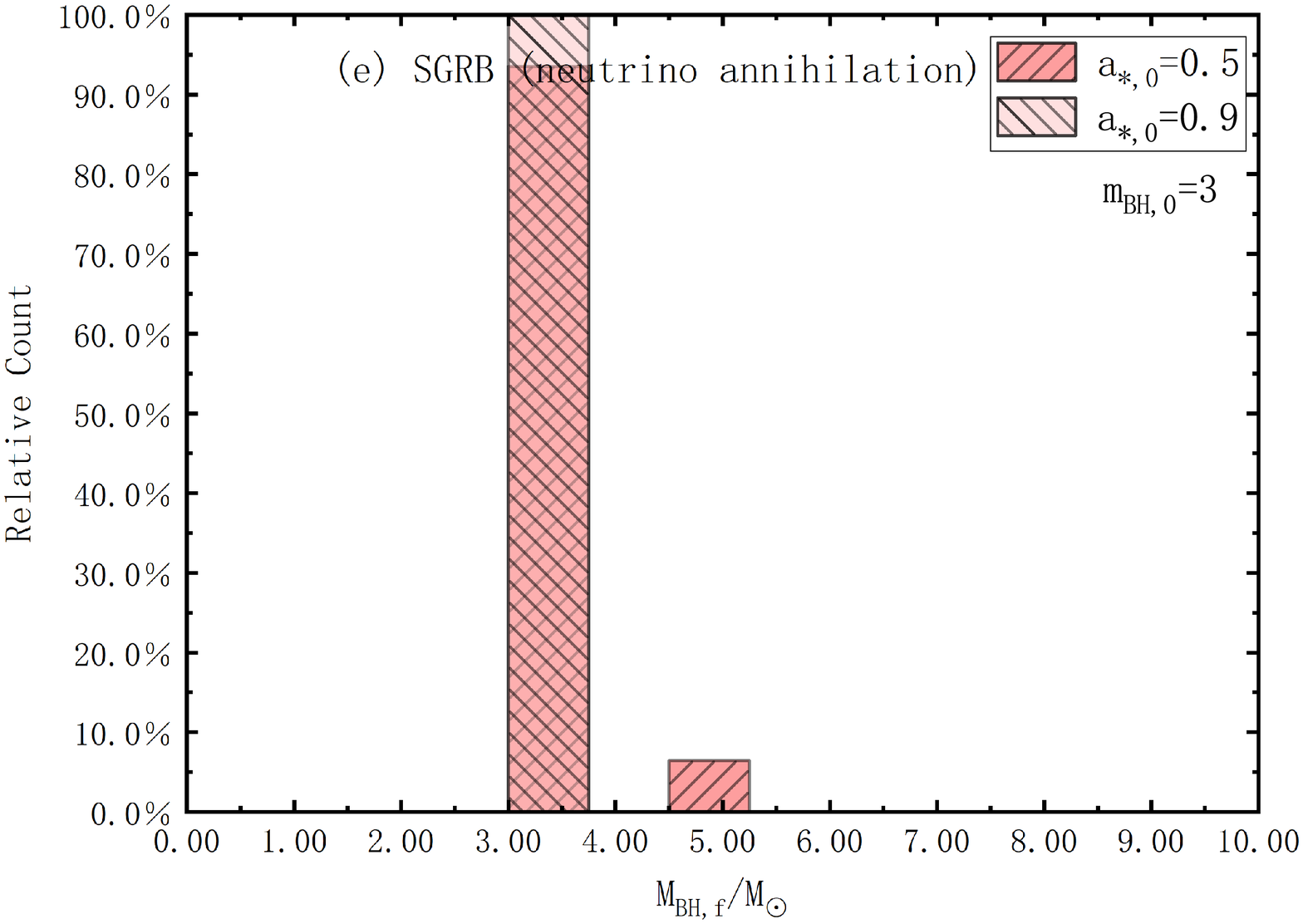}
	\includegraphics[width=0.5\linewidth]{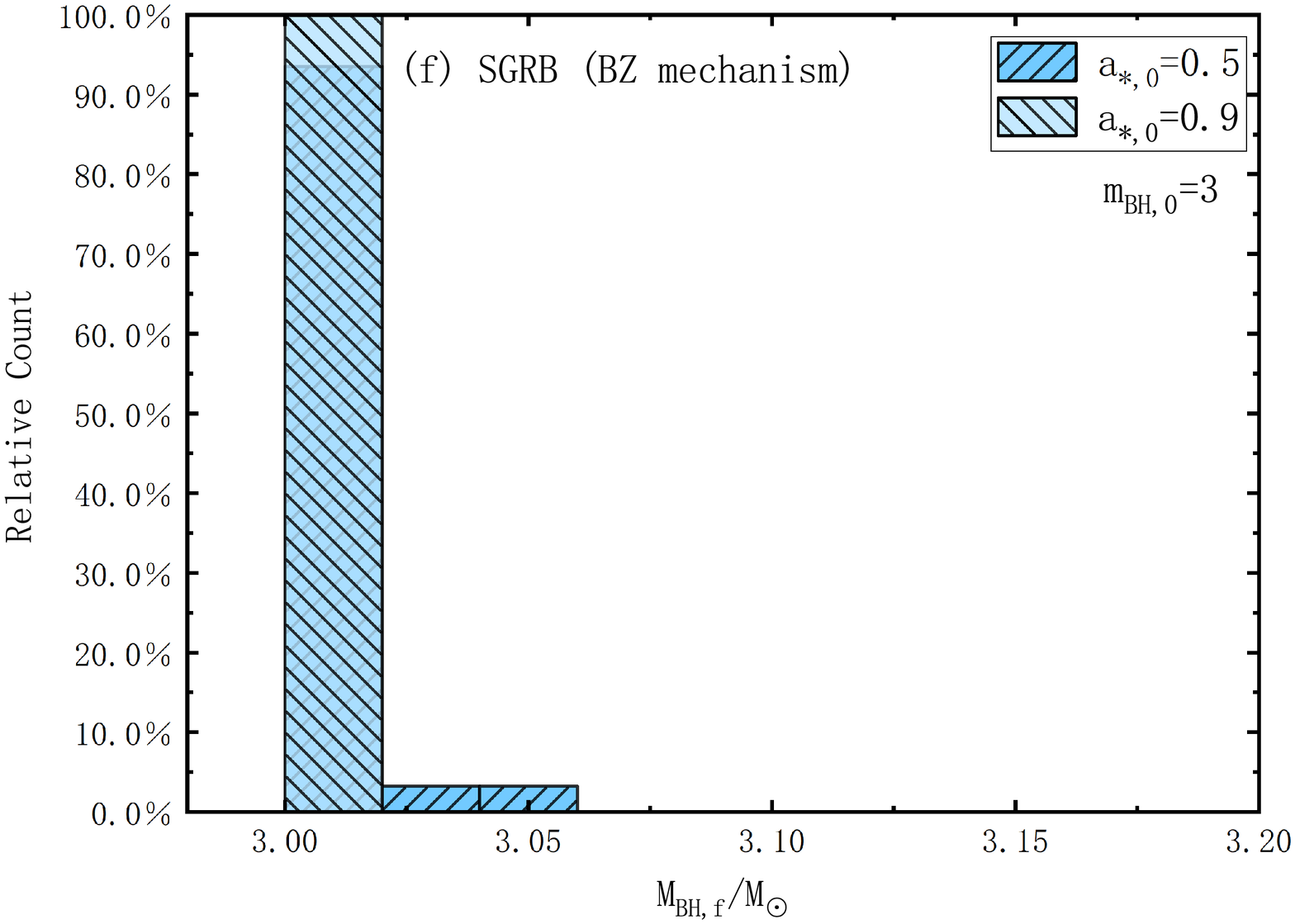}
	\caption{Distributions of the final BH masses $M_{\rm BH,f}$ for different GRB data and two different jet-launching mechanisms. The red and blue bars correspond to the neutrino annihilation process and the BZ mechanism, respectively. The initial BH mass $M_{\rm BH,0}$ is 3 $M_\odot$. The dark and light colors denote the initial BH spin $a_{*,0}$ = 0.5 and 0.9, respectively.}
	\label{fig2}
\end{figure*}

Figures \ref{fig2}(a) and \ref{fig2}(b) show the distribution of the final BH mass after the accretion phase for 14 LGRB-SN events. For the BZ mechanism, the mean BH mass growths are about $0.023 ~M_{\odot}$ for $a_{*,0}=0.5$ and about $0.003~M_{\odot}$ for $a_{*,0}=0.9$, as shown in Table \ref{tab1}. For the neutrino annihilation process, the mean BH mass growths are approximately 2.113 and $1.079~M_\odot$ for $a_{*,0}$ = 0.5 and 0.9, respectively. It is important to note that for $a_{*,0}=0.5$ under the neutrino annihilation mechanism, there are more than 40$\%$ LGRBs associated with SNe in which the BHs exceed the upper limit of the mass gap, $\sim 5~M_{\odot}$, successfully.

It should be noted that there exists inevitable competition on matter and energies between SNe and LGRBs \citep[e.g.,][]{2019ApJ...871..117S,2021ApJ...908..106L}. The SN energy depends on the initial explosion energy; however, the LGRB energy is related to the total fallback accretion mass. The typical luminosity (energy) of LGRBs associated with SNe is lower than that of LGRBs-noSNe, which is reflected by the differences in the final BH mass distribution, as shown in Figures \ref{fig2}(a-d), and the values of the mean BH mass growths, as displayed in Table \ref{tab1}. Regardless, a hydrogen envelope-deficient environment is advantageous for both events, which can reduce the stress of the energy competition. Accordingly, a massive progenitor star with powerful stellar winds would be favored as a promising progenitor of the LGRB-SN case. Once the hydrogen envelope is retained in the collapsar mode, a jet breakout should occur within hundreds of seconds, and the corresponding accretion rate is lower than the ignition of NDAFs. Then, the BZ jets monopolize the energy release. Nevertheless, for ultra-LGRBs associated with luminous SNe, such as GRB 111209A with SN 2011kl, only massive progenitors can support the ultralong activity timescale of the hyperaccretion process and violent explosion, so $>40~M_\odot$ (even $\sim 70 ~M_\odot$) progenitors are inescapably required \citep[e.g.,][]{2013ApJ...778...67N,2018ApJ...852...20L,2019ApJ...871..117S}. Moreover, if there is only a giant bump in the GRB optical afterglow and no more evidence for the existence of a CCSN associated with a GRB, one can expect that just the strong disk outflows (or winds) from hyperaccretion systems or the violent winds from magnatars could produce enough $^{56}$Ni to power the bumps without CCSN explosions \citep[e.g.,][and references therein]{2011ApJ...743..155S,2015MNRAS.451..282S,2019ApJ...871..117S}.

\subsection{LGRB-noSN case}

Figures \ref{fig2} (c) and (d) display the distribution of the final BH mass for LGRBs-noSNe. As shown in Figure \ref{fig2}(d), the BH mass has no significant increase in the BZ mechanism for $a_{*,0}=0.9$, with more than 90$\%$ of LGRBs-noSNe growing within $3.1~ M_{\odot}$. For a smaller initial BH spin parameter $a_{*,0}=0.5$, 50$\%$ of LGRBs shows growth lower than $0.1~ M_{\odot}$, while the rest can gain a mass increment between 0.1 and $0.7~ M_{\odot}$. In contrast, the mass growth in the neutrino annihilation mechanism in Figure \ref{fig2} (c) is much more promising, with many LGRBs successfully closing the lower mass gap. For the faster initial spin $a_{*,0}=0.9$, the success rate is approximately 65$\%$, while for the slower spin $a_{*,0}=0.5$, the success rate reaches 88$\%$. Additionally, there are some extreme cases in which the final BH mass increases to tens of solar masses.

For LGRBs-noSNe, the absence of associated SNe is either caused by the explosion energy being too weak or too distant to be observed. Overall, the typical energy of LGRBs-noSNe is higher than that of LGRBs associated with SNe, which implies that more massive or lower metallicity stars are the progenitors of LGRBs-noSNe or that lower explosion energy is more prevalent for their progenitors. Moreover, the mean BH mass growths are about 5.845 and 4.069 $M_\odot$ for $a_{*,0}$ = 0.5 and 0.9, respectively, as shown in Table \ref{tab1} when the neutrino annihilation process is dominant; for the BZ mechanism, the mean growths are about 0.152 and 0.026 $M_\odot$ for $a_{*,0}$ = 0.5 and 0.9, respectively.

Since the accretion rate decreases with time in the fallback accretion phase, neutrino annihilation process lasting tens of seconds should be replaced by the BZ mechanism \citep[e.g.,][]{2021ApJ...908..106L,2021MNRAS.507..431W}, and the BH mass growth might be slightly less than the above maximum values.

\begin{table*}
	\centering
	\caption{Mean BH mass growths in all cases.}
	\begin{tabular}{ccccc}
		\hline\hline
		Case  &  $a_{*,0}$ & Mechanism & Mean BH mass growth ($M_\odot$) & Figure  \\
		\hline
		LGRB-SN	&	0.5	&	Neutrino annihilation	&	2.113 	& 2(a)\\	
		LGRB-SN	&	0.9	&	Neutrino annihilation	&	1.079 	& 2(a)\\
		LGRB-SN	&	0.5	&	BZ	&	0.023 	& 2(b) \\	
		LGRB-SN	&	0.9	&	BZ	&	0.003 	& 2(b) \\
		LGRB-noSN	&	0.5	&	Neutrino annihilation	&	5.845 	& 2(c)\\
		LGRB-noSN	&	0.9	&	Neutrino annihilation	&	4.069 	& 2(c)\\
		LGRB-noSN	&	0.5	&	BZ	&	0.152 	& 2(d)\\	
		LGRB-noSN	&	0.9	&	BZ	&	0.026 	& 2(d)\\
		SGRB	&	0.5	&	Neutrino annihilation	&	0.227 	& 2(e)\\
		SGRB	&	0.9	&	Neutrino annihilation	&	0.062 	& 2(e)\\
		SGRB	&	0.5	&	BZ	&	0.0025 	& 2(f)\\	
		SGRB	&	0.9	&	BZ	&	0.0003 	& 2(f)\\
		\hline
	\end{tabular}
	\label{tab1}																
\end{table*}

\subsection{SGRB case}

Since the maximum mass of NSs constrained by the recent observations, such as GW 170817 and \emph{NICER} PSR J0030+0451, is about $2.4~M_\odot$ \citep[e.g.,][and references therein]{2020JHEAp..28...19L,2021ApJ...913...27L}, the coalescence of two NSs is hardly to produce a $>5~M_\odot$ BH but creates a BH in the lower mass gap and the following multimessenger signals. As an extreme example, in the merger of a massive NS, $\sim 2~M_\odot$, and a BH, the accretion mass could reach $\sim 0.8~M_\odot$ constrained by the SGRB extended emissions \citep[e.g.,][]{2012ApJ...760...63L}. Thus, one can expect that the BHs in the center of SGRBs cannot grow up to break through the lower mass gap if the initial accreting BH mass is set to $3~ M_{\odot}$ after mergers.

For the BH hyperaccretion system born after the merger of two compact objects, the merger ejecta hardly stops the SGRB jets, but the limited accretion matter can support no more than seconds of the central engine activity. The distributions of the final BH mass of SGRBs are shown in Figures \ref{fig2}(e) and \ref{fig2}(f). If the ultrarelativistic jet is powered by the BZ process, for both initial BH spin parameters, one can see that the growth within $\sim 0.05~ M_{\odot}$. Meanwhile, the mean BH mass growths are approximately 0.0025 and 0.0003 for $a_{*,0}=0.5$ and 0.9, respectively. If the neutrino annihilation process is dominant, although BHs with initial BH spin parameter $a_{*,0}=0.9$ still fill in the gap, there is almost no chance for BHs with the initial BH spin parameter $a_{*,0}=0.5$ to escape the gap. Furthermore, the mean BH mass growths are approximately 0.227 and 0.062 for $a_{*,0}=0.5$ and 0.9, respectively. Only two SGRBs whose central BHs grow from $3~ M_{\odot}$ to $\sim 5~ M_{\odot}$ with $a_{*,0}=0.5$, which is impossible in the merger scenario, and the BZ mechanism should be reasonable for their engines.

Thus, the NS-NS mergers can contribute a small amount of $2-5~M_{\odot}$ BHs, and the contribution of the NS-BH mergers is determined by the initial BH mass. Fortunately, the merger events are much fewer than the collapse events \citep[e.g.,][]{2004ApJ...607L..17P}, and a lower mass gap is not empty but should exist.

\section{Conclusions and discussion} \label{sec:conclusions}

In this paper, we calculated the BH mass growth using the BH hyperaccretion model and observational GRB data to investigate the contribution of hyperaccretion to lower mass gap formation. If BH hyperaccretion is considered the central engine of LGRBs, the mass of most LGRB progenitors is limited to $20-40~M_\odot$, which corresponds to the theoretical initial BH mass in the lower mass gap. As a result, one can notice that these newborn BHs could grow up and break away from the gap if the neutrino annihilation process is dominated, even just in the initial accretion phase. For the LGRB-SN case, we propose that the newborn BHs in the center of CCSNe with lower initial explosion energy have large probabilities of escaping from the gap. Of course, for progenitors without hydrogen envelopes or a low-metallicity progenitor mass larger than $40~M_\odot$, the BH might grow enough or be naturally larger than $5~M_\odot$. For the SGRB case, the BHs born in NS-NS mergers have no chance to escape from the gap, but those in NS-BH mergers are probably if the difference between the initial accreting BH mass and the upper limit of the gap is less than $\sim 1~M_\odot$.

Some X-ray plateaus and flares in GRB afterglows are believed to originate from the central engine reactivation \citep[e.g.,][]{2017NewAR..79....1L,2022ApJ...924...69Y}, and their typical luminosities are lower than $L_{\rm j}$. Nevertheless, they should further contribute to the central BH growths due to the additional accretion processes, which facilitates BHs jumping out of the lower mass gap. Of course, if the activity timescale of the GRB central engine is much longer than $T_{90}$ \citep[e.g.,][]{2014ApJ...787...66Z,2017NewAR..79....1L}, the BHs should get more opportunities to grow up to $> 5~M_\odot$.

According to the standard external shock model \citep[e.g.,][]{Zhang2018}, the jet half-opening angle $\theta_{\mathrm{j}}$ can be estimated by the observed jet breaks in X-ray afterglows. Unfortunately, it is difficult to get the accurate values for both LGRBs and SGRBs since the absence of jet break observations, so the lower limits of $\theta_{\mathrm{j}}$ are widely used \citep[e.g.,][]{2015ApJ...806...58L,2017JHEAp..13....1Y}. As shown by Equation (\ref{eq6}), the larger $\theta_{\mathrm{j}}$ should amplify $L_{\mathrm{j}}$ then be beneficial to the BHs escaping from the gap.

The lower mass gap problem arises from the observations of X-ray binaries and are widely discussed since the first direct detection of gravitational waves (GWs). LGRBs are considered to originate from the collapses of the single stars. Actually, many stars are in binaries or even multiple systems \citep[e.g.,][]{2013ARA&A..51..269D}, and LGRBs can also be triggered in the phases of the common envelopes \citep[e.g.,][]{2000ApJ...532..540A} or the binary systems \citep[e.g.,][and references therein]{2021ApJ...921....2Z}. Even so, it is almost impossible to result in the X-ray binaries after these explosions as well as SGRBs, which means that GRBs and X-ray binaries might be the differently plausible channels in the step of the first-generation mergers \citep[e.g.,][]{2021NatAs...5..749G}. The future detections on the electromagnetic radiation accompanied with neutrinos and GWs from the massive collapsars might give more clues on the BH mass growth undergoing explosions \citep[e.g.,][]{2021MNRAS.507..431W}. Moreover, after the hierarchical mergers of stellar-mass BHs in their history \citep[e.g.,][and reference therein]{2021NatAs...5..749G}, most of BHs in the lower mass gap might be eliminated at high redshift and disappeared in the local universe. Based on the above discussion, we consider that the lower mass gap in the mass distribution of compact objects exists but not empty, and its formation involves the GRB contribution.

Mass outflows might occur in the BH hyperaccretion system, which will participate in nucleosynthesis to power kilonovae or SNe in merger or collapsar scenarios \citep[e.g.,][]{2011ApJ...743..155S,2018MNRAS.477.2173S,2019ApJ...871..117S,2021ApJ...920....5L}. Once the effects of outflows are considered, based on the GRB data, the above results on the values of the BH mass growths should be their lower limits because the disk outflows will weaken the neutrino annihilation luminosity accumulated from the whole disk or the strength of the magnetic fields binding in the disk. Whatever, the hyperaccretion mode is the only way to significantly influence the BH evolution and the mass distribution of compact objects for the single stars.

The ultrarelativistic jets are distinctly unobservable if they are out of sight or choked in the envelopes or circumstances. Therefore, the GRB sample in our work can only partially represent the contribution of the hyperaccretion process to the lower mass gap. Nevertheless, jets, disks, mergers, and explosions are still strong sources of neutrinos and GWs \citep[e.g.,][]{2016PhRvD..93l3004L,2017ApJ...850...30L,2019ApJ...878..142W,2020ApJ...889...73W}, and one can expect further joint multimessenger observations to describe the shape of the lower mass gap.

\acknowledgments
We thank Shu-Yu Hu for the helpful discussion. This work was supported by the National Natural Science Foundation of China under grants 12173031 and 11822304 and the science research grants from the China Manned Space Project with No. CMS-CSST-2021-B11. H.M.Q. acknowledges support from the Undergraduate Innovation Program of Xiamen University.

\end{document}